\documentstyle[12pt]{article}
\begin{document}
\textheight 9in
\topmargin -0.5in
\begin{titlepage}

\begin{flushright} IFUM/432/FT\\
                   SISSA/14/94/EP\\

                   December 93
\end{flushright}
\vfill
\begin{center}
{\large\bf Exact renormalization group equation and decoupling
in  quantum field theory}   \\
\vskip 5.mm
{\bf Luciano Girardello$^1$}\\
Dipartimento di Fisica, Universita' di Milano\\
via Celoria 16, I-20133 Milano, Italy\\and I.N.F.N. sezione di
Milano\\[0.3cm]

{\bf Alberto Zaffaroni$^2$} \\
SISSA - International School for Advanced Studies
\\via Beirut 2, I-341000 Trieste, Italy
\\and I.N.F.N. sezione di trieste

\end{center}
\vfill
\begin{abstract}
New proves of decoupling of massive fields in several quantum field
theories are derived in the effective Lagrangian approach
based on Wilson renormalization group. In the most interesting
case of gauge theories with spontaneous symmetry breaking,
the approach, combined with the quantum action principle, leads to
a rather simple proof to all orders.
\end{abstract}
\vfill
\vfill
\begin{quote}
\vskip 2mm
\hrule width 5.cm
\vskip 1.mm

{\small\small
\noindent $^1$ e-mail address: girardello@vaxmi.mi.infn.it\\
\noindent $^2$ e-mail address: zaffa@tsmi19.sissa.it\\
\noindent $^*$ Work supported in part by Ministero dell'Universit\'a e
della Ricerca Scientifica e Tecnologica}
\normalsize

\end{quote}
\end{titlepage}

\section{Introduction}
\label{sec:intro}
Most of the conceptual understanding of the renormalization
of quantum field theories
via the Wilson renormalization group can be directly retrieved from
the works of the same Wilson~\cite{Wil-Kog}. Having obtained in dependence
of a cut-off $\Lambda$ a flow of physically equivalent lagrangians (i.e.
with the same Green functions), he realizes that the renormalizability
of the theory relies on the existence of an infrared attracting surface,
parametrized by the relevant coupling (i.e. the renormalizable interaction,
in the language of quantum field theory). To implement this approach Wilson
proposed to trade  the functional integration over the high momentum modes
for a
functional differential equation governing the effective Lagrangian L. This
equation, which translates the indipendence of the partition function
on the floating cut-off $\Lambda$, gives rise to an infinite system of
first order differential equations in $\Lambda$ for the coefficients
(vertices) of the expansion of L in power series of the fields of the
theory.\par
In spite of the appeal of this more  intuitive approach no progress was made
in this direction. The real investigation of the structure of the field
theory has been
pursued along the old and complex diagrammatic way; the proof of the
perturbative
renormalizability of the theory and of the most important theorems in quantum
field theory (the Wilson operator expansion, the action principle and so on)
is  worked out in the BHPZ framework, for example.\par
 It is only few years ago that Polchinski~\cite{Pol}
has shown how to exploit such a differential formulation. A recursive,
perturbative method of solution, within an inductive approach,
allowed
him to derive appropriate bounds on the effective vertices of the 4-D
Euclidean scalar theory and then to prove the existence of an infrared
three dimensional critical surface,i.e. renormalizability. Remarkably,
this method avoids completely the hard problems of the standard
approach, above all that  of the  overlapping divergences.
This has prompted a renewed interest in the Wilson- Polchinski
approach : extensions to QED, to SU(2) Yang-Mills
theory~\cite{keller}~\cite{marche}  and
a thorough reformulation of the whole approach~\cite{becchi}.
The technical simplicity of the method makes it worth to explore
applications to general questions as well as to specific issues of QFT
and to come up to a more intuitive and deep understanding
 (~\cite{ellwanger}
and references therein).\par
The purpose of this article is to discuss the problem of the
decoupling of a field of very large mass in this more general
contest. The method seems, indeed, particularly suitable for the
discussion of such a problem since it is  mainly a low energy affair.
All the results we present exist already  in the literature, but their
proof at all order in perturbation theory is only given in a heuristic
way or simply sketched, making appeal  to the cumbersome BHP  for the
more rigorous  demonstrations. We want to show how to derive
rigorously all these results in the Polchinski scheme, without facing graphical
analysis or other standard complications.\par
The well understood state of the art for the decoupling problem is the
following:
if we study a physical problem in which one of the particle has a very large
mass M we say that it decouples if at low energy its effect is suppressed
by powers of M. In the analysis of the  physics of the light particles,
 it enters through an internal heavy
line in the Feynman graphs: decoupling is expected if these
diagrams are suppressed; more precisely we require that their effects, if
not O(1/M), can be absorbed into finite renormalization of masses and
coupling constants. For theories without spontaneous symmetry breaking, the
decoupling of heavy particles was demonstrated by Appelquist and
Carazzone~\cite{app}. In theories with spontaneous symmetry breaking, however,
we have decoupling only when we increase a mass by increasing a dimensional
parameter (for example a vacuum expectation value). We will discuss all
these topics more diffusely in the next sections.\par
This paper is organized as follows:
in section 2 we sketch the Polchinski renormalization scheme, giving
a self-contained proof of renormalizability of a scalar field.
In section 3,4 we recall the general facts about decoupling and
reobtain the Appelquist-Carazzone theorem; the part of the analysis
related to dimensional arguments is very simple, for the
gauge problem we choose to work in the BRS approach, invoking the powerful
action principle. We give also a brief discussion of the general structure of
the first order correction to decoupling. In section 5 we discuss
the physical more interesting case of decoupling in theory with
spontaneous symmetry breaking, and we present a largely model
independent analysis.\par
\section{The Polchinski scheme}
\label{sec:pol}
In this section we want to discuss the exact renormalization group
equations, the renormalization scheme based on them, and the
perturbative proof of the ultraviolet finiteness of the euclidean
scalar field.~\cite{Pol}\par
In order to compute any Green function of a field theory one needs a
 regularization procedure of the ultraviolet divergences.
Let's consider a scalar field in four euclidean dimensions, with kinetic part
\begin{equation}
{1\over 2}\int {d^4p\over (2\pi)^4}
\phi(p)(p^2+m^2)\phi(-p)K^{-1}(p^2/\Lambda^2)
\end{equation}
where $K(p^2/\Lambda^2)$ is a $C^\infty$ function which vanishes rapidly at
infinity and assumes the value 1 for $p^2<\Lambda^2$. We have
regularized the theory by eliminating from the propagator
the modes greater than a certain cut-off.\par
{}From a complementary point of view, we may think of  $\Lambda$ not merely as
a
regulator to avoid divergences but also as a physical scale at which to compute
physical quantities. In the Wilson approach to the renormalization group we
start from an initial lagrangian at the scale $\Lambda_0$ and define
an
effective lagrangian obtained by integrating out frequencies between
$\Lambda_0$ and $\Lambda$.
We can associate with a given
 interaction lagrangian $L(\phi,\Lambda)$ the partition
function
\begin{eqnarray}
Z(J,\Lambda) = \int[d\phi]exp [&-&{1\over 2}\int{dp\over(2\pi)^4}
\phi(p)K^{-1}(p^2/\Lambda^2)(p^2 + m^2)
\phi(-p)\nonumber \\ &+& L(\phi,\Lambda )
+ \int{dp\over (2\pi)^4}J(p)\phi(-p) ]
\end{eqnarray}
where we integrate over the remaining (low energies) modes; it's obvious
that $Z(\Lambda)$ contains all the physical informations of the theory
and is independent of the scale $\Lambda$.\par
More generally, without reference to the Wilson point of view, we suppose
to have an interaction lagrangian at a certain scale $\Lambda$ and the
associated partition function.
We want to probe the physics under a certain physical scale $\Lambda_R$, so
we take the source of the following form: J(p)=0 for $p^2>\Lambda^2_R$
(for an alternative point of view in which $\Lambda_R$ is zero,
see~\cite{marche}).\par
We obtain a flow in the space of the theories by imposing that the partition
function does not change when we vary the cut-off. It is easy
to verify~\cite{Pol} that, if the interaction lagrangian satisfies the
following evolution equation,
\begin{equation}
\Lambda\partial_{\Lambda}L(\phi,\Lambda) = -{1\over2}\int dp (2\pi)^4
\Lambda{\partial_\Lambda K(p^2/\Lambda^2)\over p^2+m^2}\left\{{\delta^2L\over
\delta\phi(p)\delta\phi(-p)} + {\delta L\over\delta\phi(p)} {\delta L
\over\delta\phi(-p)} \right\}
\label{eq:flow}
\end{equation}
it follows that $\Lambda\partial_{\Lambda}Z(J,\Lambda)=0$. In this way we
obtain
a traiectory in the space of lagrangians which have the same
Green functions for momenta below $\Lambda_R$, and
hence have the same physical
content at low energy. As shown in the paper of Polchinski, this
flow is strongly attracted by an infrared three-dimensional surface,
parametrized by the coefficients of the relevant (i.e. power
counting renormalizable)
interaction.
If we expand the lagrangian in the following manner (assuming for the theory
the symmetry $\phi\rightarrow -\phi$, preserved by the flow)
\begin{equation}
L(\phi,\Lambda)=\sum_{m=1}^{\infty}{1\over(2m)!}\int {dp_1 ... dp_{2m}\over
(2\pi)^{8m-4}}L_{2m}(p_1,...,p_{2m};\Lambda)\delta^4(p_1+ ... +p_{2m})
\phi(p_1) ... \phi(p_{2m})
\label{eq:lag}
\end{equation}
and define also the dimensionless vertices of the theory
\begin{equation}
L_{2m}(p_1, ... ,p_{2m};\Lambda)=A_{2m}(p_1, ...
,p_{2m};\Lambda)/\Lambda^{2m-4}
\end{equation}
we obtain the flow equation in the following form
\begin{eqnarray}
&&(\Lambda\partial_{\Lambda}+4-2m)A_{2m}(p_1, ... ,p_{2m};\Lambda)=\nonumber\\
&-&{1 \over2}
\sum_{l=1}^m Q(P,m,\Lambda)A_{2l}(-P,p_1, ... ,p_{2l-1};\Lambda)\times
\nonumber\\
&&A_{2m-2l+2}(P,p_{2l}, ... ,p_{2m};\Lambda)|_{(P=p_1+....+p_{2l-1})}+Perm.
\nonumber\\
&-&{1\over2}\int {dp\over(2\pi\Lambda)^4}Q(p,m,\Lambda)A_{2m+2}(p,-p,p_1, ...
,p_{2m};\Lambda)
\label{eq:vertices}
\end{eqnarray}
where
\begin{equation}
Q(p,\Lambda,m^2)={1\over (p^2+m^2)}\Lambda^3{\partial\over \partial
\Lambda}K(p^2/\Lambda^2)
\end{equation}
and where we sum over all the possible
 combinations of 2l-1 impulses $(p_1, ...
,p_{2l-1})$ out of $(p_1, ... ,p_{2m})$.\par
It might seem somewhat irrelevant to study this evolution equation,
for we already know the solution in terms of an effective lagrangian
in which we have integrated out the high frequencies down to $\Lambda$;
this corresponds to compute a first functional integral over the high modes.
More precisely~\cite{keller}~\cite{marche}~\cite{becchi},
if we start at the scale
$\Lambda_0$ from the lagrangian $S(\Lambda_0)$, it is easy
to show that the evolved lagrangian
can be written as the functional generator of the connected and amputated
Feynman diagrams corresponding to the vertices of $S(\Lambda_0)$ and
the following propagator
\begin{equation}
\Delta(\Lambda,p) = {K(p^2/\Lambda_0^2) - K(p^2/\Lambda^2)\over p^2 + m^2}
\end{equation}
So the vertices
$L_{2m}$ are exactly the Feynman graphs of the theory described. However,
this
diagrammatic representation is noteworthy but technically cumbersome.
Arguments which refers to it necessarily suffer from all the
complications of the graphical approach, while the direct use of the
equation overcomes all these problems. We will refer to this explicit
solution when the intuition will need it and when the understanding
 of certain aspects will be enhanced from this point of view.\par
The use of the equation~(\ref{eq:flow}) in the study of the non perturbative
aspects of quantum field theory is still unclear, while the perturbative
solution has been well studied~\cite{Pol}.We will outline, in the
following, its main features with emphasis on
the technical aspects which are needed for our problem.\par
Up to now, we have the equations for a flow of lagrangians, parametrized
by the cut-off $\Lambda$, which give the same answer for the Green functions;
the flow starts from the initial lagrangian at $\Lambda_0$ and ends at a
certain physical scale $\Lambda_R$. The simultaneously presence of three
different cut-off might confuse the reader: as it will be soon clear,
$\Lambda_0$ must be considered the regulating cut-off to be removed,
$\Lambda_R$ the energy scale at which we impose the renormalization conditions,
and $\Lambda$ is only an interpolating parameter, an independent variable in
a differential equation.\par
We define for the general lagrangian~(\ref{eq:lag}) the "relevant"
parameters:
\begin{equation}
\rho_1(\Lambda)=-L_2(0,0;\Lambda);\hskip 4pt
\rho_2(\Lambda)=-{1\over8}{\delta^2
\over\delta^2p}L_2(p,-p;\Lambda)|_{p=0};\hskip 4pt \\
\rho_3(\Lambda)=-L_4(0,0,0,0;\Lambda)
\end{equation}
they are exactly the terms of positive or null mass dimension in the
lagrangian (i.e. the coefficients of $\phi^2, (\partial\phi)^2, \phi^4)$;
note that we don't distinguish between marginal and relevant
terms in Wilson sense.\par
We want now to construct a flow of lagrangians which converges
to the renormalized $\lambda\phi^4$ theory. We
impose that, at the initial scale $\Lambda_0$, the lagrangian consists
only of relevant terms, so we have a {\it bare} lagrangian
of the form:
\begin{equation}
\int d^4x\left( {1\over2}(\partial\phi)^2+{m^2\over2}\phi^2+{\rho_1
(\Lambda_0)\over2}\phi^2+{\rho_2(\Lambda_0)\over2}(\partial\phi)^2+
{\rho_3(\Lambda_0)\over4!}\phi^4\right)
\end{equation}
We want to stress the interpretation of this expression as
a bare lagrangian regularized with a cut-off $\Lambda_0$: so we have
\begin{equation}
\rho_1(\Lambda_0)= \delta m^2,\qquad \rho_2(\Lambda_0)=Z-1,\qquad
\rho_3(\Lambda_0)
=\lambda_0Z^2
\end{equation}
We impose now as a second condition (at all effects a {\it renormalization}
condition) that at low scale $\Lambda_R$ the relevant parameters assume
the values:
\begin{equation}
\rho_1(\Lambda_R;\Lambda_0,\rho_0)=\rho_2(\Lambda_R;\Lambda_0,\rho_0)=0,
\qquad \rho_3(\Lambda_R;\Lambda_0,\rho_0)=\lambda_R
\end{equation}
We seek for a flow which satisfies both the conditions in perturbation theory,
i.e. when we consider the vertex $A_{2m}$ expanded in power of the
{\it renormalized} coupling $\lambda_R$; whereas a non
perturbative solution
is not guaranteed, a perturbative solution always exists.
The geometric spirit of the renormalizability theorem consists in what
follows: when we increase the cut-off $\Lambda_0$ imposing
the previous conditions, the $\Lambda_R$ lagrangian is forced to
converge
to the critical surface, with an error
of the order $(\Lambda_R/\Lambda_0)^2$.
The complete proof of this statement can be found
in~\cite{Pol}~\cite{becchi}.
 Since
the theory at the point $(0,0,\Lambda_R)$ on the critical
surface is finite (like every theory with a cut-off) and since
the lagrangians on the same traiectory have the same low energy Green
functions,
we conclude that we can take the limit $\Lambda_0 \rightarrow \infty$ obtaining
a finite theory.\par
This method can be used to prove the UV-finiteness of all the power
counting renormalizable quantum field theories; we will sketch the proof in the
simple case of $\lambda\phi^4$, but, since the
general argument is essentially based on dimensional considerations,
extensions to more complicated theories is straightforward but
for chiral theories. The general analysis
must be done perturbatively by expanding all the vertices in formal
power series of $\hbar$; the only difference in the proof is that the
induction must be done increasing the number of external legs in the
vertices and paying attention to the correct $\hbar$ power of the
parameters.
We want only to note that the equation~(\ref{eq:flow}) involves a not
well-defined functional integral, so we have to introduce an infrared cut-off
which reduce the problem to a finite numbers of Fourier modes.
 The problem is easily by-passed
observing that the equation for the vertices is perturbatively well-defined
in every case. The thermodynamic limit defines
a consistent quantum field theory.
For a more precise discussion of all these points we refer to
reference~\cite{becchi} and we will pursue a more qualitative
discussion following Polchinski's presentation.\par
We want now to give a review of the proof of the finiteness of
 the vertices $A_{2m}$
in the limit $\Lambda_0\rightarrow\infty$.\par
We will simplify the argument by taking a function $K(p^2/\Lambda^2)$ which
vanishes for $p\geq2\Lambda$. We introduce the norm
\begin{equation}
\|f(p_1, ...,p_m)\|=\max_{p_i^2\leq4\Lambda^2}|f(p_1, ... ,p_m;\Lambda)|
\end{equation}
it is straightforward, then, to verify that
\begin{eqnarray}
&&\int{dp\over(2\pi)^2}|Q(p,\Lambda,m)|<C\Lambda^4\nonumber\\
&&\left\|{\partial^n\over\partial
p^n}Q(p,\Lambda,m)\right\|<D_n{1\over\Lambda^4}
\label{eq:Q}
\end{eqnarray}
with C and $D_n$ constants independent of m.\par
In the perturbative theory we have the great simplification that
$A_{2m}^{(0)}=0$ and, at order r in $\lambda_R$, $A_{2m}^{(r)}=0$
for $m>r+1$ .
This is easily understood if we observe that at order r a diagram has at most
2r+2 external legs. This fact suggests an induction approach to the
perturbative
solution to the equations~(\ref{eq:vertices}): $A_{2m}^{(s)}$ can be computed
when we know the vertices at the lower perturbative orders and those
of the
same order with more external legs; since they are in a finite number,
for the previous observation, this double inductive procedure will
lead to
the complete solution of the equation. In the following of this
paper all the proofs, also without explicit declaration, will be by
induction.\par
{\bf LEMMA 1} At order r in $\lambda_R$
\begin{equation}
\|\partial^pA_{2m}^{(r)}(p_1,...,p_{2m};\Lambda)\|\leq{1\over\Lambda^p}
P(ln{\Lambda\over\Lambda_R})
\label{eq:lemma}
\end{equation}
with P a non negative coefficients polinomial and the
derivation with respect with the momenta is of order p. The precise form
of the derivative operators is not necessary
for the argument.\par
Proof: notice, first of all,
\begin{equation}
\int {dx\over x}{P^{(m)}(lnx)\over x^q}=-{P_1^{(m)}(lnx)\over x^q},\qquad q>0
\end{equation}
(with P, $P_1$... positive coefficient polynomials of order m), hence:
\begin{equation}
\int_{\Lambda}^{\Lambda_0}{d\Lambda'\over\Lambda'}P^{(m)}\left(ln{\Lambda'
\over\Lambda_R}\right)\left({\Lambda\over\Lambda'}\right)^q \leq \tilde
P^{(m)}
\left(ln{\Lambda\over\Lambda_R}\right),\qquad  q>0
\label{eq:int}
\end{equation}
We assume that the equation~(\ref{eq:lemma})
is true for $A_{2n}^{(s)}$ with $n>m$ up to the perturbative
 order s-1 and proceed downward in m. The first step of
the double induction is satisfied because $A_{2m}^{(0)} = 0$ and
$A_{2m}^{(s)} = 0$
for $m >s + 1$. By considering the norm of the perturbative version of
equation~(\ref{eq:vertices}) we have
\begin{equation}
\|(\Lambda\partial_\Lambda + 4 - 2m) A_{2m}^{(s)}(\Lambda)\| \leq \sum_{l=1}
^m \sum_{t=1}^{s-1}\|A_{2l}^{(t)}\|\|A_{2m-2l+2}^{(s-t)}\| + \|A_{2m=2}^{(s)}\|
\end{equation}
where we have used equation~(\ref{eq:Q}) and the fact that the function Q
forces
the momenta in the range $[\Lambda,2\Lambda]$.
In the following we shall always ignore positive
coefficients in these bounds.
The same equation for derived vertices
can be deduced by using again eq.~(\ref{eq:Q}), where it's needed.
We stress that the only effect of a momentum derivative
is a factor $1/\Lambda$.
 By the induction hypothesis
\begin{equation}
\|(\Lambda\partial_\Lambda + 4 - 2m) \partial^p A_{2m}^{(s)}(\Lambda)\| \leq
{1\over\Lambda^p}P(ln\Lambda/\Lambda_R)
\end{equation}
For $m \geq 3, A_{2m}^{(s)}(\Lambda_0) = 0$. Integrating the previous formula
between $\Lambda$ and $\Lambda_0$, we obtain
\begin{equation}
\|\partial^pA_{2m}^{(s)}(\Lambda)\| \leq {1\over\Lambda^p}
\int_{\Lambda}^{\Lambda_0}{d\Lambda'\over\Lambda'}P\left(ln{\Lambda'\over
\Lambda_R}\right)\left({\Lambda\over\Lambda'}\right)^{p+2m-4}
\leq {1\over\Lambda^p}\tilde P\left(ln{\Lambda\over\Lambda_R}\right)
\end{equation}
where we can use eq.~(\ref{eq:int}) because $p+2m-4>0$. For the same
reason eq.~(\ref{eq:lemma}) is valid for $m=2, p\geq1$. The case p=0
\begin{equation}
|\Lambda\partial_\Lambda A_4^{(s)}(0,0,0,0;\Lambda)| \leq
P(ln\Lambda/\Lambda_R)
\end{equation}
must be integrated from $\Lambda_R$ because we don't know $A_4(0;\Lambda_0)$,
which doesn't vanish,
\begin{eqnarray}
&&|A_4^{(s)}(0,;\Lambda)| \leq |A_4^{(s)}(0;\Lambda_R)|
+ |A_4^{(s)}(0;\Lambda) -
A_4^{(s)}(0;\Lambda_R)| = \delta^{s1}\nonumber\\
&& + \int_{\Lambda_R}^{\Lambda}
{d\Lambda'\over\Lambda'}P\left(ln{\Lambda'\over\Lambda_R}\right)
\leq \delta^{s1} + P\left(ln{\Lambda\over\Lambda_R}\right)\int_{\Lambda_R}
^{\Lambda}{d\Lambda'\over\Lambda'} = \tilde P\left(ln{\Lambda\over\Lambda_R}
\right)
\end{eqnarray}
Since $A_4^{(s)}(p_1, ... ,p_4;\Lambda)$ can be reconstructed via Taylor
theorem
from $A_4^{(s)}(0;\Lambda)$ and $\partial^2A_4^{(s)}(p-1, ... ,p-4;\Lambda)$,
(\ref{eq:lemma}) is proved for $A_4^{(s)}$. \par
It is straightforward to complete the proof in the case m=1
following the prescriptions to
integrate downward
the irrelevant terms $(m=1, p>2)$ from $\Lambda_0$ where we know that
the vertex $A_2$ is zero, and upward the relevant terms (p=0,2 at zero
momentum)
from $\Lambda_R$ where we know the corresponding values (the renormalization
conditions). The result for the whole vertex is obtained via a Taylor
expansion based on
these informations.\par
\par
 We refer, once more, to the original paper of Polchinski
for the detailed discussion of the approaching
to the critical surface; here we want only
to indicate how to obtain from these bounds the proof of
$\Lambda_0$ (i.e. ultraviolet) finiteness of our theory.\par
The very same bounds seem to indicate that at a finite scale the
vertices are $\Lambda_0$ finite, because the coefficients in the polinomials
are positive numbers independent of any dimensional parameter of the
theory. More precisely, the explicit solution of the flow equation can
be obtained by induction from the following formula for $m \geq 3$,
\begin{eqnarray}
L_{2m}^{(s)}(\Lambda) &=& -{1\over2}\int_{\Lambda_0}^{\Lambda}d\Lambda'
\left( \sum_{l=1}^m \sum_{t=1}^{s-1} {dK/d\Lambda\over P^2+m^2}L_{2l}^{(t)}
(\Lambda')L_{2m-2l+2}^{(s-t)}(\Lambda')\right)_{P=p_1+...+p_{2l-1}}
\nonumber\\
&+& Perm.
-{1\over2}\int_{\Lambda_0}^{\Lambda}d\Lambda'\int {dp\over(2\pi)^4}
{dK/d\Lambda\over p^2+m^2}L_{2m+2}^{(s)}(\Lambda')
\end{eqnarray}
We suppose, as induction hypothesis, that for $\Lambda_0\rightarrow\infty$
$L_{2p}^{(t)}, t<s$ and $L_{2p}^{(s)}, p>m$ exist. It's straightforward to
bound uniformly in $\Lambda_0$ the previous integrands with convergent
 expressions, deducing the $\Lambda_0$-finitness of $L_{2m}^{(s)}$.\par
With $\Lambda_o$ finite we can obtain the previous formula obtaining the
finiteness of $\partial^pL_{2m}$ for $m=2, p\geq1, m=1, p\geq2$. The
result
in the relevant cases is again easier since the integration is over
the finite range $[\Lambda_R,\Lambda]$. The use of Taylor theorem
completes
the proof.
\section{Decoupling for theory without spontaneous symmetry
breaking}
\label{sec:dec}
If we work at a scale of energy such that some particles of our theory cannot
be
produced, are we able  to detect experimentally their existence? In
general, for
renormalizable theories without spontaneous symmetry breaking
a decoupling theorem can be
proved: when the heavy mass M is much larger than all the light
masses and scales
 of observation, we can compute all the scattering
amplitudes of light particles
from an effective light theory obtained with a (light) parameter
redefinition. Since the physical quantities have to be fixed from
the experiments, such redefinition is undetectable. More precisely,
if M is much greater than all the light masses, the external
momenta and the renormalization scale, we have that the total light
1PI functions
can be computed from an effective
lagrangian to $O(1/M^{\alpha})$ precision, i.e.
\begin{equation}
\Gamma_n(p_i,g,m,M) = Z^{-n/2}\Gamma_n(p_i,g^*,m^*) + O(1/M^{\alpha})
\end{equation}
where $g^*, m^*$ represent the redefined set of light couplings and masses.
The explicit form of decoupling is very sensitive to the kind of
renormalization
we use. Naively, we expect that every graph  with an internal heavy line is
suppressed by inverse powers of M: this is not true, of course, due to
 the existence
of renormalization and counterterms which can depend in
an unexpected way on M.
What is true is that all these effects can be reabsorbed in coupling
constant, mass and wave function redefinitions~\cite{col}. The case in which
these graphs are in fact driven to zero and such renormalization of
the parameters is not needed is called {\it manifest} decoupling. The
typical
scheme in which the decoupling is manifest is the BHPZ~\cite{col}
{}~\cite{zimm};
for example if a graph is already finite with ultraviolet degree $-\delta$
($\delta>0$) we expect that it behaves as $1/M^{\delta}$. On the other hand,
if the graph diverges, the effect of the renormalization is to subtract the
appropriate order of the
 Taylor expansion, leading to the same expression derivated as many
times as to leave it with a negative UV degree, to which
we apply the previous reasoning.\par
Recalling the representation of vertices as Green functions given in
section 2 and the form of the renormalization conditions, we recognize
many
analogies of this
 scheme with the BHPZ scheme, when we consider our vertices as
the amplitudes which must be renormalized in the latter one.
An explicit computation of the lower orders vertices by the flow equations
confirms this point of view: it's very simple, for example, to recognize
the expression of the two or four point functions in terms of Feynman
graphs renormalized by subtraction of the Taylor series.
We want to show now in a simple model that
in the Polchinski scheme decoupling exists,
as expected, and it is manifest.
\par
We work in a theory of two
scalar fields interacting with a quartic potential,
\begin{equation}
L = (1/2)((\partial\phi)^2 + m^2\phi^2) +  (1/2)((\partial\chi)^2 + M^2\chi^2)
+ \lambda_R\phi^4 + g_R\chi^4 + k_R\phi^2\chi^2
\end{equation}
Deriving the obvious generalization of
eq.~(\ref{eq:vertices}) with respect to M and denoting
\begin{equation}
W_{2m,2n} = M {\partial A_{2m,2n}\over \partial M}
\end{equation}
where the index m refers to light and n to heavy fields, we obtain
\begin{eqnarray}
&&\left(\Lambda{\partial\over \partial\Lambda} + 4 - 2m - 2n \right)W_{2m,2n}
= -\sum_{l=1}^m \sum_{p=0}^n {\Lambda^3 dK/d\Lambda\over P^2+m^2}
A_{2l,2m}W_{2m-2l+2,2n-2p}\nonumber\\
&&- {1\over2}\int {dp\over (2\pi)^4\Lambda}{dK/d\Lambda\over p^2+m^2}
W_{2m+2,2n}\nonumber\\
&&-\sum_{l=0}^m\sum_{p=1}^n {\Lambda^3dK/d\Lambda\over P^2+M^2} A_{2l,2p}
W_{2m-2l,2n-2p+2} - {1\over2}\int {dp\over(2\pi)^4\Lambda}{dK/d\Lambda\over
p^2+M^2}W_{2m,2n+2}\nonumber\\
&&+\sum_{l=0}^m \sum_{p=1}^n {\Lambda^3dK/d\Lambda\over (P^2+M^2)^2}
M^2A_{2l,2p}A_{2m-2l,2n-2p+2} + \int {dq\over (2\pi)^4\Lambda}
{dK/d\Lambda\over (q^2+M^2)^2}M^2A_{2m,2n+2}\nonumber\\
&&\null
\end{eqnarray}
We want to show that
\begin{equation}
\|\partial^pW_{2m,2n}^{(r)}(\Lambda)\| \leq {1\over\Lambda^P}\left(
{\Lambda\over M}\right)^{2-\epsilon}P\left(ln{\Lambda\over\Lambda_R}\right),
\qquad for \hskip 4pt\epsilon\in(0,1]
\label{eq:w}
\end{equation}
where the coefficients of the polinomials can depend on $\epsilon$ and
one cannot take
the limit $\epsilon\rightarrow0$.\par
We work again by induction, assuming eq.~(\ref{eq:w}) true
at order $t<s$ and at
order s for a total number of legs greater than $2(m+n)$. Obviously,
$W_{2m,2n}^{(0)}=0$ and $W_{2m,2n}^{(s)}=0$ for $m+n>s+1$.\par
For the M propagator we will make use of the bound
\begin{equation}
\left|\int {dp\over (2\pi)^4\Lambda}{dK/d\Lambda\over p^2+M^2}\right|
\leq D(\epsilon)\left(\Lambda\over M\right)^{2-\epsilon},\qquad \epsilon
\in [0,1]
\end{equation}

Using eq.~(\ref{eq:Q}) for the homogeneous part of the equation,
the bounds~(\ref{eq:w}) for the inomogeneous one, and the bounds on
the A vertices we
have proved in the previous section, we obtain:
\begin{equation}
\|(\Lambda\partial_\Lambda+4-2m-2n)\partial^pW_{2m,2n}^{(s)}(\Lambda)\|
\leq {1\over\Lambda^p}\left({\Lambda\over M}\right)^{2-\epsilon}P\left(ln
{\Lambda\over\Lambda_R}\right)
\end{equation}
For $m+n\geq3$, $W_{2m,2n}^{(s)}(\Lambda_0)=0$ ; the upward integration
gives:
\begin{eqnarray}
&&\|\partial^pW_{2m,2n}^{(s)}
(\Lambda)\|\leq {1\over\Lambda^p}\left({\Lambda\over M}
\right)^{2-\epsilon}\int_{\Lambda}^{\Lambda_0}{d\Lambda'\over\Lambda'}
P\left({\Lambda\over\Lambda'}\right)^{p+2m+2n-6+\epsilon}\nonumber\\
&&\leq {1\over\Lambda^p}\left({\Lambda\over M}\right)^{2-\epsilon}\tilde P
\left(ln{\Lambda\over\Lambda_R}\right)
\end{eqnarray}
We note that for this kind of proof the parameter $\epsilon$ is unavoidable.
The rest of the proof is identical to that of the lemma: we integrate upward
 the relevant terms, the value of the W's at this scale being zero.
We stress that the proof relies on the fact that
the inomogeneous term in the
equation controls
 the behaviour of the vertices and that the initial conditions
don't destroy it: the rest of the demonstration is analogous
to the one for the $A_{2m}$ vertices.\par
We find
\begin{eqnarray}
&&\|A_{2m,2n}^{(s)}(\Lambda,\Lambda_0;m,M) -
A_{2m,2n}^{(s)}(\Lambda,\Lambda_0;m,M')\| =\\
&&\|\int^{M'}_{M}{dM\over M}W_{2m,2n}^{(s)}(\Lambda,\Lambda_0;m,M)\|
\leq P(ln{\Lambda\over \Lambda_R})(2-\epsilon)
\left|\left({\Lambda\over M'}\right)^{2-\epsilon}
 - \left({\Lambda\over M}\right)^{2-\epsilon}\right|\nonumber
\end{eqnarray}
For the uniformity of the bounds, we can take the limit $\Lambda_0
\rightarrow\infty$. Chauchy criterium guarantees also the existence
of the $M\rightarrow\infty$.
\par We can
 compare our bound with the results of J.Ambjorn~\cite
{amb} for the amplitudes of Feynman graphs in the BHPZ scheme when M is sent
to $\infty$, which read
\begin{equation}
|F(p_i,m,M)|\leq C(\epsilon){1\over M^{2\nu-\epsilon}},
\qquad for \hskip 4pt\epsilon\in(0,1)
\end{equation}
where $\nu \geq 1$. The result is obtained via the parametric representation
of Feynman graph, with much more work.\par
We must again prove that $A_{2m,0}^{(M\rightarrow\infty)}$ agrees with
the result of the theory with only light fields. $A_{2m,0}$ satisfies the
equation
\begin{eqnarray}
&&(\Lambda\partial_\Lambda+4-2m)A_{2m,0}^{(s)} = -{1\over2}\sum_{l=1}^m
\sum_{t=1}^{s-1}Q(P,m;\Lambda)A_{2l,0}^tA_{2m-2l+2,0}^{(s-t)}\nonumber\\
&&-{1\over2}\int{dp\over(2\pi\lambda)^4}Q(p,m;\Lambda)A_{2m+2,0}^{(s)}
-{1\over2}\int{dp\over(2\pi\Lambda)^4}Q(p,M;\Lambda)A_{2m,2}^{(s)}
\end{eqnarray}
The equation can be integrated to give the explicit solution. Using the bounds
proved above
and the same tricks of the previous arguments, it is easy to show that
that we can take the limit $M\rightarrow\infty$ under the sign of integrals
and that, in this limit, the third integral vanishes; so the $A_{2m,0}^{M
\rightarrow\infty}$ satisfies the same equations of $A_{2m}$ (when we
write
 only one index in A we refer to the light theory) and, since, at the lowest
 order, they coincide, they coincide at all orders (the solution is
uniquely determined by recursion from the initial conditions).\par
We can also investigate the structure of $O(1/M^2)$
corrections to light fields
vertices. Working in electrodynamics, Kazama and Yao~\cite{kazama}
find for the corrections the following structure
\begin{equation}
\Gamma_n(g,G,m,M)=\Gamma_n(g,m)+M^{-2}\sum_i C_i(g,G,ln(m/M))\Gamma_n(O_i,g,m)
\label{eq:corr}
\end{equation}
where g is the light coupling constant, G the heavy one, $O_i$ are integrated
local
composite operators of dimension $\leq6$ and $C_i$ are universal coefficients,
calculated via a kind of Callan-Symanzik equation.
We are interested in deducing this structure from the flow equation as a
valid alternative to the BHPZ combinatorial formulas used by these authors.
\par First of all we must review some facts on the renormalization of
composite operators.
\par
We introduce a source $\epsilon$
for the composite operator and expand the lagrangian
in the following way;
\begin{equation}
S_0(\phi,\Lambda)+\epsilon S_1(\phi,\Lambda)+\epsilon^2S_2(\phi,\Lambda)+...
\end{equation}
where $S_1$ is the running operator insertion, $S_2$ is needed for
double insertions and so on.\par
The equation of flow for the whole S implies for $S_1$ a linearized equation,
which, for the vertices $O_{2m}$, (analogous to $L_{2m}$) reads
schematically:
\begin{equation}
\partial_{\Lambda}O_{2m}=\sum L_{2l}O_{2m-2l+2} + \int O_{2m+2}
\end{equation}
$S_1$ is determined at low energy by its relevant parameters.
For more details
we refer to~\cite{warr,becchi}. For example, for the renormalization of
$\phi^6$, we specify the following renormalization conditions
\begin{equation}
Q_6(0,...,0,q;\Lambda_R)=1 \hskip 4pt and \hskip 4pt
\eta_i(\Lambda_R)=0;\hskip 4pt at
\hskip 4pt\Lambda_0
\hskip 4pt only\hskip 4pt the \hskip 4pt\eta_i \not=0
\end{equation}
where $Q_{2m}$ are the adimensional counterparts of $O_{2m}$ and $\eta_i$ are
the operator of dimension $\leq6$ allowed by symmetry. We see that the
counterterms must be chosen among
all the operator of dimension $\leq6$,
i.e, as we know, they mix under renormalization. We finally note that now
$O_{2m}^{(s)},m>s+3$ and $O_{2m}^{(0)}\not=0$.\par
To exploit the counterpart of Kazama-Yao result we consider again two scalar
fields coupled by a quartic interaction of the form $k_R\phi^2\chi^2$.
Note that we can always
choose composite operator $O_i$ (of dimension 6) and dimensionless
functions $C_i(m,M,\Lambda_R,\lambda_R,k_R)$ such that the quantities
\begin{equation}
C_{2m}(p_i;\Lambda)=-{1\over 2}M^3\partial_ML_{2m,0}(p_i;\Lambda)
-\sum_iC_i(m,M,\Lambda_R,\lambda_R,k_R)O_{2m}^i(p_i;0;\Lambda)
\end{equation}
vanish on the relevant parameters (dim $\leq6$) at $\Lambda_R$. For
example
\begin{equation}
C_{\phi^6}=-{1/2}M^3\partial_ML_{6,0}(0;\Lambda_R)
\end{equation}
We want to show that $|C_{2m}^{(s)}(\Lambda_R)|\rightarrow0 \hskip 4pt
for \hskip 4pt M\rightarrow
\infty$ which is the analogous of eq.~(\ref{eq:corr}).
Notice that we recover the result of Kazama and Yao for the vertices of
our theory and not for the real Green functions. The analogy doesn't
break for we can consistently regard the $\Lambda_R$ vertices as the
connected Green functions of our theory, renormalized according to
the prescription of the flow equations; as we have already noted,
in this way one produces a sort of BHPZ regularization.\par
The rest of this section is devoted to the proof of this statement;
the reader not interested in the technical details can skip the proof
and go to the next section. We want only to stress that the final
structure is dictated by dimensional analysis.
The form of the insertion of
composite operators is dictated by the canonical dimension of the quantity
considered: the factor $1/M^2$ raises the reference dimension to 6,
and in order
to use the standard arguments we must introduce a quantity which has
compatible initial data: the only way
is to subtract a suitable 6-dimension operator.
\par First of all notice that $L_{2m,0}^{(s,1)}(\Lambda_R)=0$ ( we denote with
the pair (s,1) the perturbative order in $\lambda_R$ and $k_R$ separately);
this
follows from the graphycal representation of the vertex: containing only one
$k_R$
the diagram is necessarily divergent and it vanishes by renormalization.
So $C_i$ are at least of second order in $k_R$.
Moreover it is easy to show that
\begin{eqnarray}
\|L_{2m,0}^{(p,q)}(\Lambda)\|\leq (\Lambda/M)^{2-\epsilon}(1/\Lambda^{2m-4})
P(ln\Lambda/\Lambda_R),\qquad &q&\geq1\nonumber\\
\|L_{2m,2}^{(p,q)}\|\leq (\Lambda/M)^{2-\epsilon}(1/\Lambda^{2m-2})
P(ln\Lambda/\Lambda_R),\qquad &q&>1
\label{eq:20}
\end{eqnarray}
so they vanish when $M\rightarrow\infty$; of course
it is essential for the proof
that they have null initial condition.
\par We already know that $C_i$ and so $C_{2m}(\Lambda_R)$ are of second
order in $k_R$. Let's show that
\begin{equation}
\|C_{2m}^{(p,q)}\|\leq\left({\Lambda\over M}\right)^{2-\epsilon}
{1\over \Lambda^{2m-6}}
P\left(ln{\Lambda\over\Lambda_R}\right)\left({M\over\Lambda_R}\right)^{\epsilon
'}
, q\geq1, M>\Lambda_R, \epsilon,\epsilon'\in(0,1)
\end{equation}
Due to the uniformity of these bounds, the limit $\Lambda_0\rightarrow\infty$
is then immediate.\par
The $C_{2m}$ satisfy
\begin{eqnarray}
&&{\partial\over\partial\Lambda}C_{2m}=-\sum {dK/d\Lambda\over P^2+m^2}
L_{2l,0}C_{2m-2l+2} -{1\over2}\int{dp\over(2\pi)^4}{dK/d\Lambda\over
p^2+m^2}C_{2m+2}\nonumber\\
&&-{1\over2}\int{dp\over(2\pi)^4}{dK/d\Lambda\over
p^2+M^2}M^3\partial_ML_{2m,2}
-{1\over2}\int{dp\over(2\pi)^4}{dK/d\Lambda\over (p^2+M^2)^2}M^4L_{2m,2}
\nonumber\\
&&+\sum_iC_i{dK/d\Lambda\over P^2+m^2}(L_{2l,0}-L_{2l})O_{2m-2l+2}^i
\label{eq.44}
\end{eqnarray}
Now $C_{2m}^{(p,0)}=0$ because $C_i=O$ and
$\partial_ML^{(p.0)}_{2m,0}=0$
(it doesn't contain heavy lines). In the first term on the right hand side
there can appear a term $LC^{(t,1)}$, not covered by the induction hypothesis;
however L will be of the form $L^{( ,\geq1)}$ for which we the
bounds~(\ref{eq:20}) hold.
But we need also the weaker bound
\begin{equation}
\|\partial^pC_{2m}^{(t,1)}\|\leq \left({1\over\Lambda^{2m-6}}\right)
P\left(ln{\Lambda\over\Lambda_R}\right){1\over\Lambda^p}
\left({M\over\Lambda}\right)^
{\epsilon}
\end{equation}
which follows from the definition of $C_{2m}$ and from the bounds over the
M-derivative of $L_{2m}$.\par
We use $|C_i|\leq D(m/\Lambda_R)^{\epsilon '}$,
\begin{eqnarray}
\|L_{2l,0}-L_{2l}\|=\|\int_M^{\infty}{dM\over M}M\partial_M L_{2l,0}\|
&\leq& C(\epsilon){P(ln\Lambda/\Lambda_R)\over\Lambda^{2l-4}}\left({\Lambda
\over M}\right)^{2-\epsilon}\nonumber\\
&&\|\partial^pO_{2m}^{i (s)}\|\leq {1\over\Lambda^p}{P(ln\Lambda/\Lambda_R)
\over \Lambda^{2m-6}}
\end{eqnarray}
we obtain
\begin{equation}
\|\partial_{\Lambda}\partial^pC_{2m}^{(p,q)}\|\leq {1\over\Lambda}
P\left(ln{\Lambda\over\Lambda_R}\right){1\over\Lambda^{2m-6}}
\left({\Lambda\over M}\right)^{2-\epsilon}\left({M\over\Lambda_R}\right)^
{\epsilon'},\qquad q>1, M>\Lambda_R
\end{equation}
the first step of induction is obtained from the same
equation~\ref{eq.44}.
 The rest of
the proof is straightforward: for $m\geq4$
(irrelevant terms for the quantities considered)
we integrate downward
from $\Lambda_0$, and in the relevant case upward from $\Lambda_R$,
where we have suitably chosen the $C_i$ in such a way to have null initial
conditions, instead of a complicated and unknown M dependence.

\section{Gauge theories}
\label{sec:gauge}
We have proved so far that the theory of interacting scalar fields exhibits
decoupling; the generalization to arbitrary power-counting renormalizable
theories with global symmetry groups is straightforward, because the
proof is essentially based on the dimensional analysis.
The problems in gauge theories is the same of their renormalizability:
it is simple to construct an UV finite quantum theory but in general if
we don't use an invariant regularization or clever renormalization conditions
the final theory will have lost the
explicit symmetry embodied in the Ward identities,
which we consider the natural implementation of the concept of symmetry
in the quantum theory.\par
A higher derivative
gauge invariant regularization which allows the use of Polchinski
arguments is presented
in the work of Warr~\cite{warr}, whereas  a
dimensional regularization for the flow is still lacking. As usual the
regularized theory satisfies the Ward identities and what we have to
prove is that it's possible to specify the renormalization conditions
(i.e. the low energy data) in such a way that the cut-off removal is
innocuous. We prefer to use the combination of the quantum action
principle~\cite{action} or BRS approach~\cite{piguet},
which we briefly review here.\par
In the gauge fixed lagrangian
\begin{equation}
L=-{1\over 4g^4}trF_{\mu\nu}F^{\mu\nu}+L_{MAT}-{1\over 2\alpha}
tr(\partial_{\mu}A^{\mu})^2+{1\over\alpha}trcM\overline c
\end{equation}
where $M(.)=\partial^2-i\partial_{\mu}[A^{\mu},.]$ we have lost
the gauge invariance but we have gained the invariance
under the BRS transformations
\begin{eqnarray}
\Delta A _{\mu}&=&\partial_{\mu}\overline c+i[\overline c,A_{\mu}]
\nonumber\\
\delta \phi&=&i\overline c_iT^i\phi\nonumber\\
\Delta\overline c=i\overline c\overline c
&=&-(1/2)f_{ijk}\overline c^i\overline c^k\nonumber\\
\Delta c&=&\partial A
\end{eqnarray}
These are nihilpotent transformations so that also the lagrangian in which we
introduce the inert sources for the composite fields
$\Delta A,\Delta\phi,\Delta\overline c$
\begin{equation}
L' =-{1\over 4g^4}trF_{\mu\nu}F^{\mu\nu}+L_{MAT}-{1\over 2\alpha}
tr(\partial_{\mu}A^{\mu})^2+{1\over\alpha}trcM\overline c+
tr(\rho^{\mu}\Delta A_{\mu}+U\Delta\overline c)+Y^a\Delta\phi_a
\end{equation}
is BRS invariant. To renormalize a gauge theory means
to construct a quantum extension $\Gamma$ of the lagrangian
(effective action) which has the gauge symmetry expressed by the Slavnov-Taylor
identities $\Delta\Gamma (A,\phi ,c,\overline c;\rho ,U,Y)=0$. At the classical
level this identity specifies completely the lagrangian up to a redefinition
of parameters and wavefunctions~\cite{piguet}.\par
$\Delta^2\Gamma=0$ implies that
$\Gamma$ depends only on $\eta_{\mu}=\rho_{\mu}-(1/\alpha)\partial_{\mu}c$.
Defining
\begin{equation}
\overline\Gamma(A,\phi ,\overline c;\eta ,Y,U)=\Gamma+{1\over 2\alpha}
\int dx(\partial A^i)^2
\end{equation}
the Slavnov-Taylor identities read
\begin{equation}
\Delta(\Gamma)={1\over2}B_{\overline\Gamma}\overline\Gamma=0
\end{equation}
where $B_{\gamma}$ is the linear operator
\begin{equation}
B_{\gamma}=\int dx\left\{{\partial\gamma\over\partial\eta_{\mu}^i}
{\partial\over\partial A^{\mu}_i} + {\partial\gamma\over\partial A_i^{\mu}}
{\partial\over\partial\eta^i_{\mu}} + {\partial\gamma\over\partial Y^a}
{\partial\over\partial\phi_a} + {\partial\gamma\over\partial\phi_a}
{\partial\over\partial Y^a} +
 {\partial\gamma\over\partial U^i}
{\partial\over\partial\overline c_i} +
{\partial\gamma\over\partial\overline c}{\partial\over\partial U^i}\right\}
\end{equation}
which satisfies
\begin{eqnarray}
B_{\gamma}B_{\gamma}\gamma=0\\
B_{\gamma}B_{\gamma}=0 \qquad if \qquad B_{\gamma}\gamma=0
\end{eqnarray}
At order zero $\Gamma_0$ is nothing but the classical
action $\int d^4xL'$, and the linear
operator $b=B_{\overline\Gamma_0}$ satisfies
\begin{equation}
b\overline\Gamma_0=0, \qquad b^2=0
\end{equation}
On the fields
$A,\phi,\overline c$ it corresponds to the BRS transformation.\par
We begin by regularizing the theory by a cut-off; the power of the
BRS approach is that it is independent of the type of
regularization. The Polchinski scheme produces an UV finite theory
which, of course, has the global symmetry but not the gauge one
and which exhibits decoupling.\par
To fix the ideas, let us consider a Yang-Mills theory
minimally coupled with massive scalar fields, and let us send the mass
to $\infty$.\par
The powerful action principle, which can be proved in the
Polchinski scheme~\cite{becchi}  (see the next section),
 states that the possible
violations of the Ward identities are integrated
insertions of a local operator of
dimension $\leq5$ and of Faddeev-Popov number 1
\begin{equation}
\Delta(\Gamma)=\Delta\Gamma
\end{equation}
If we can reabsorb the anomaly $\Delta$ in local counterterms
in the action we have a renormalizable gauge invariant
quantum theory. Suppose, by induction, that $\Delta$ vanishes up
to $\hbar^{n-1}$ order,
\begin{equation}
\Delta(\Gamma) = {1\over 2}B_{\overline\Gamma}\overline\Gamma =
 \Delta + O(\hbar\Delta)
\end{equation}
 $\Delta$ is a globally invariant field polynomial, c independent and
 moreover is a b-closed
functional.
In fact to $\hbar^n$ order
\begin{equation}
0=B_{\overline\Gamma}B_{\overline\Gamma}\overline\Gamma=b\Delta+
O(\hbar\Delta)
\end{equation}
The purpose is to classify all the possible anomalies;
because $b^2=0$ all $\Delta=b\Delta'$ are possible anomalies;
 $\Delta'$
are of dimension 4 and FP number 0 so they can be obviously reabsorbed in
the lagrangian by defining $L'=L-\Delta'$,
\begin{equation}
\Delta(\Gamma')=\Delta(\Gamma)-\Delta(\Delta')+O(\hbar^{n+1})=O(\hbar^{n+1})
\end{equation}
The theory is now renormalizable and anomaly free to this order in $\hbar$.
We can repeat the reasoning order by order.
If the theory is not chiral, one can prove, in a purely algebraic way, that
b-exact functionals are all the possible solutions of $b\tilde\Delta=0$;
 it is obviously a cohomological problem.~\cite{piguet}\par
We have so found a quantum gauge extension of our Yang-Mills
theory; obviouslywe have to change the renormalization conditions
 by adding new counterterms. Wehave again a freedom: it is easy
 to verify that adding to the lagrangianthe general solution
 of $b\Delta=0$ in the space of four dimensionalfunctionals, we don't
 break the symmetry; this freedom corresponds to reparametrize the initial
bare classical lagrangianand we can use it to impose some renormalization
 conditions.\par
{}From this point of view the extension of the decoupling theorem to
gauge theory does not present problems. We work by induction in $\hbar$,
supposing to have a symmetric and decoupled theory; we renormalize it
in the Polchinski scheme at order $\hbar^n$ and we obtain a decoupled theory
but with an anomaly $\Delta(\Gamma)=\tilde a+
O(\hbar a)$.
Looking at the left hand side, we see that $\tilde a$
 has M-finite coefficients which
coincide with the anomaly of the decoupled theory if we take the limit
$M\rightarrow\infty$; when reabsorbed in the action they have again the same
property, establishing the symmetry at this order without violating
manifest decoupling.\par
We have so proved the Appelquist-Carazzone theorem. Obviously if we want
to consider the question of the effective lagrangian in a different
renormalization scheme we must renormalize coupling constants and
wavefuctions.
\section{Theory with spontaneous symmetry breaking}
\label{sec:sbs}
So far we have proved that a general renormalizable theory without
spontaneous symmetry breaking exhibits decoupling. In the case of
SSB we have two ways of increasing a mass, either with a dimensional parameter
(typically a vacuum expectation value), in
which case we have decoupling~\cite{kazama2}, or with a dimensionless
one (a coupling constant) in which case the decoupling theorem fails.
A typical example of the last situation is the limit
for $\lambda\rightarrow\infty$ of the
linear $\sigma$-model
which gives rise to the non renormalizable non linear $\sigma$-
model. For definiteness, we consider the
following lagrangian
\begin{equation}
{1\over2}(\partial h)^2 + {1\over2}(\partial H)^2 + {m^2\over2}h^2 +
{M^2\over2}H^2 + \lambda_1h^4 + \lambda_2H^4 + \lambda_3(h\times H)^2 +
\lambda_4h^2H^2
\end{equation}
with h and H in the vectorial representation of SO(3). The group is
completely broken in the ortogonal directions (0,v,0) and (V,0,0)~\cite{gild}
and the parameter v and V take the role of the two masses m and M as free
parameters; we want to send V to $\infty$. As already
mentioned, we expect decoupling.\par
To study the decoupling in a scalar theory,
it has been proposed~\cite{das} a variation of the $\overline{MS}$ scheme
 where heavy graphs,
too large in power of M, are eliminated ad hoc by local counterterms.
This scheme is probably only suitable for the study of the decoupling
problems. A complete proof of decoupling also in gauge
theories has been
provided ~\cite{das2}. Another proof
 is presented by Kazama and Yao~\cite{kazama2}
working in the traditional BHPZ scheme.\par
We first note that in the Polchinski approach we have an immediate
problem: all the bounds we have given were derived by assuming that every
dimensional parameter
appearing in the lagrangian is lower than the scale $\Lambda_R$, otherwise
the bounds on the vertices will be violated even at the classical level.
 However the breaking
introduces
 trilinear terms of the form $M\chi\eta^2$ (where
$h=v+\eta, H=V+\chi$). We have to modify the renormalization rules;
a simple way is
to employ a solution similar to the Chang-Das one~\cite{das}.\par
If we are able to renormalize the theory and obtain vertices of order $M^n$,
where n is the number of
heavy legs, we conclude, by power counting, that also the Green functions
(which we can evaluate at the scale $\Lambda_R$ where the integration of the
Feynman diagram are finite)
with n heavy lines are of order $M^n$. In particular, this implies the
result that the Green functions with only external light fields are M-finite.
\par
These considerations and the lagrangian form in the fields $\eta,\chi$
suggest the bound
\begin{equation}
\|L_{m,n}\| \leq \left({M\over\Lambda}\right)^n\Lambda^{4-m-n}
P\left(ln{\Lambda\over\Lambda_R}\right)
\end{equation}
We use hereafter an induction in $\hbar$ so, at fixed order
in $\hbar$, we must increase
the number of external legs. At tree level these bounds are verified.
Assuming their validity at order n-1 we discover easily that, at order n,
they are again verified for the irrelevant parameters.
But for the relevant ones we obtain the bound:
\begin{equation}
\|\partial_{\Lambda}L_{m,n}\| \leq P\left({M\over\Lambda}\right)^{(n)}
\Lambda^{4-m-n}P\left(ln{\Lambda\over\Lambda_R}\right)
\end{equation}
to be integrated between $\Lambda$ and $\Lambda_R$; here P denotes a polynomial
of maximum degree n.
The terms which have total negative powers of $\Lambda$
invalidate the usual argument: we are forced
 to treat them as irrelevant
and to integrate them downward
from $\Lambda_0$. In other words, we explicitly choose
 the renormalization conditions (i.e. the low energy data)
to cancel the unwanted terms, with a strong analogy with the method of
Chang and Das. However, in our approach, it is
obvious that we introduce only counterterms of dimension 4, while the
previous authors
must refer to a modified version of the forest formula and the relative
proof.
\par
We have obtained a decoupled theory. What's about gauge theories?
We consider the same model with SO(3) gauge group~\cite{kazama2}.
The use of t'Hooft gauge fixing $(\partial A^a - (\eta,T^av) -(\chi,T^aV))^2$
distinguishes between heavy and low sector. We don't precise further
the model, because our argument is largely detail-independent.\par
 Evaluating the anomalous Slavnov identities
at the scale $\Lambda_R$, we learn
that the general term in the anomaly with n
heavy fields is $O(M^{n+1})$. For our purpose this is not enough;
luckily, we can employ again
the action
principle. \par
The quantum action principle~\cite{action} states, roughly speaking,
that a classical symmetry can be violated at the quantum level, i.e.
for the generating functionals Z and $\Gamma$,
at most by a local insertion of specified dimension.
It was first proved
in the BHPZ framework with a lot of technical difficulties, and then for
other renormalization scheme. In the Polchinski approach, the proof is very
simple~\cite{becchi},
 just because the method is a sort a legitimation of the heuristic
functional integral arguments.\par
Working again with a volume cut-off (before the thermodynamic limit)
we can make the change of variables in the partition function
\begin{equation}
\phi_p\rightarrow\phi_p+\sum_q\epsilon_{p-q}P_{0,q}(\phi)
\end{equation}
after the introduction of the source $\eta$ for the composite operator P
in the action $L_0$, obtaining
\begin{equation}
\sum_{p,q}\epsilon_pJ_{-q}\partial_{\eta_{-p-q}}Z = \sum_q\epsilon_q\Delta_qZ
\end{equation}
where
\begin{equation}
\Delta_p = \sum_p(\phi_{-p}P(p)+\partial_{\phi_p}L_0)P_{o,p-q} -
\partial_{\phi_p}P_{0,p-q}
\end{equation}
We can repeat this change of variable at each scale $\Lambda$. The first
hand side of the equation for Z is UV finite so also the insertion operator
$\Delta$ is
finite. This is the expression of the quantum action principle.
The first non vanishing order of $\Delta$  is the anomaly a which appears in
$\Delta(\Gamma)=a+O(\hbar a)$; despite the appareance, it is finite and
scale independent. If we
substitute the vertices bounds in the explicit expression
of $\Delta$ we find terms of the form
\begin{equation}
\left({M\over\Lambda}\right)^n\Lambda^{4-m-n}\times polynomials
\end{equation}
Evaluating this expression at the
scale $\Lambda_0$ we find that only terms with
power of M lower than 6 can survive in the anomaly.\par
So we have only few terms to check. They have dimension 5 and FP
number 1. Possible  terms of the type $M^4\overline c$ in the anomaly
violate the bound $M^{n+1}$; $M^3\overline c(.)$ produces, once reabsorbed,
 linearterms, which we can
eliminate by adjusting V and v to give the right vacuum.
The remaining terms must be checked model by model; having a number of free
parameters to adjust order by order, we can control them. For example,
in the specific model indicated, we have some further symmetries. The theory
and the BRS transformation are invariant under $\chi\rightarrow-\chi,
M\rightarrow-M,K\rightarrow-K$. It exists also an index symmetry~\cite{kazama2}
which prevents many vertices. So, by a simple check, we are left only with
the light fields quadratic terms. The normalization conditions are
surely enough to
eliminate them. We notice also that for models with less
symmetries,  the Ward identities will
relate some of the unknown terms to the already bounded ones,
leaving a few number of independent terms, which we can control with
the renormalization conditions.\par
\vskip 1truecm
It is a pleasure to acknowledge C. Becchi for enlightening discussions.
We would also like to thank M. Raciti e F. Riva for useful
discussions.

\end{document}